\documentstyle [12pt] {article}
\title{THERE IS NEITHER CLASSICAL BUG WITH A SUPERLUMINAL SHADOW
NOR QUANTUM ABSOLUTE COLLAPSE NOR (SUBQUANTUM) SUPERLUMINAL HIDDEN
VARIABLE}
\author{Vladan Pankovi\'c$^{\ast,\star}$, Milan
Predojevi\'c $^{\ast,\star}$,Miodrag Krmar$^{\ast}$
Milan Radovanovi\'c $^{\star}$ \\
$^{\ast}$Department of Physics,\\Faculty of Natural Sciences and
Mathematics,\\21000 Novi Sad,Trg Dositeja Obradovi\'ca 4.,Serbia
and Montenegro,
\\vladanp@gimnazija-indjija.edu.yu\\
$^{\star}$ Gimnazija,22320 Indjija,Trg Slobode 1a,Serbia and
Montenegro}

\date {}
\begin{document}
\maketitle

\vspace {1cm}

PACS number:    03.65.Ta , 03.65.Ud , 03.30.+p

\vspace{2cm}

\begin {abstract}

In this work we analyse critically Griffiths's example of the
classical superluminal motion of a bug shadow. Griffiths considers
that this example is conceptually very close to quantum
nonlocality or superluminality,i.e. quantum breaking of the famous
Bell inequality. Or, generally, he suggests implicitly an absolute
asymmetric duality (subluminality vs. superluminality) principle
in any fundamental physical theory.It, he hopes, can be used for a
natural interpretation of the quantum mechanics too. But we
explain that such Griffiths's interpretation retires implicitly
but significantly from usual, Copenhagen interpretation of the
standard quantum mechanical formalism. Within Copenhagen
interpretation basic complementarity principle represents, in
fact, a dynamical symmetry principle (including its spontaneous
breaking, i.e. effective hiding by measurement). Similarly, in
other fundamental physical theories  instead of Griffiths's
absolute asymmetric duality principle there is a dynamical
symmetry (including its spontaneous breaking, i.e. effective
hiding in some of these theories) principle. Finally, we show that
Griffiths's example of the bug shadow superluminal motion is
definitely incorrect (it sharply contradicts the remarkable
Roemer's determination of the speed of light by coming late of
Jupiter's first moon shadow).
\end {abstract}

\newpage

In this work we shall analyze critically presentations,
discussions and implicitly suggested original solutions of some
significant conceptual problems of the quantum mechanics
foundation given in the significant book of David Griffiths {\it
Introduction to Quantum Mechanics} [1] .Precisely, we shall
analyze Griffiths's suggestion on the conceptual analogy between
quantum mechanics (quantum nonlocality or superluminality, i.e.
quantum breaking of the famous Bell inequality) and classical
mechanics (classical (nonquantum) superluminal motion of a bug
shadow along a sufficiently distant screen). This analogy,
Griffiths hopes, implies the existence of a {\it absolute
asymmetric duality}(causal subluminality, i.e. locality vs.
acausal superluminality, i.e. nonlocality) principle not only in
the quantum mechanics but in practically all fundamental physical
theories (classical mechanics, theory of relativity, quantum
mechanics and quantum field theory). Further we shall compare such
Griffiths's absolute asymmetric duality  principle with
"relativistic and symmetrical" suppositions and statements of the
usual, Copenhagen (Bohr-Heisenberg's) interpretation of the
standard quantum mechanical formalism. It will  be  demonstrated
that Griffiths's and Copenhagen conceptual view points are
principally different. Moreover, it will be demonstrated that
conceptual basis of practically all fundamental physical theories
(classical mechanics, theory of relativity, quantum mechanics and
quantum field theory) represents instead of the Griffiths's
absolute asymmetric duality principle a "relativistic" dynamical
symmetry (including its spontaneous breaking, i.e. effective
hiding in some of these theories) principle. Finally, we shall
prove that Griffiths's example of the classical superluminal bug
shadow motion is definitely inconsistent. It is not hard to see
that given example contradicts sharply the remarkable Roemer's
determination of the speed of light by coming late of  Jupiter's
first moon shadow. Subtle distinctions between absolute asymmetric
(Griffiths's and other) and "relativistic and symmetric"
(Copenhagen) interpretations of the quantum mechanics, which are
presented here, can be very interesting for the students for a
better understanding of the basical concepts in the modern
physics.

In the afterward of his above-mentioned book David Griffiths
presents simply and clearly famous Einstein-Podolsky-Rosen
paradox, EPR [2] (that states that quantum mechanics is {\it
consistent} but {\it incomplete} so that it must be completed with
some (subquantum) hidden variables), Bell inequality [3]  (that
states that {\it if }quantum mechanics is {\it incomplete then} it
is {\it necessarily inconsistent}) and takes notice on Bell
inequality breaking by real experiments [4]. These works,
Griffiths concludes, imply the so-called {\it quantum nonlocality,
superluminality} or (by Einstein's words [2]) "spooky
action-at-a-distance". Or, generally, it implies, in Griffiths's
opinion, a principal absolute asymmetric duality (causal
subluminal dynamics vs. accusal superluminal influences) not only
in quantum mechanics but in all other fundamental physical
theories too.

However this conclusion is not quite accurate even if  the term
"quantum nonlocality" is in a general use. Quite correctly
speaking, according to its standard formalism [5] including its
usual, Copenhagen (Bohr-Heisenberg's) interpretation [6],[7], that
are in an excellent agreement with real experiments [4], quantum
mechanics is both, {\it complete and subluminal}. It forbids not
only superluminal energy or information transfer but {\it any}
superluminal action, i.e. influence. Namely, according to standard
quantum mechanical formalism and Copenhagen  interpretation
seeming superluminal actions represent noting more than a
consequence of an {\it incorrect use} of this formalism.
Especially, collapse of the wave function by measurement, as one
of the most fundamental standard quantum mechanical concepts,
represents, as it has often been suggested by Bohr (even if Bohr
does not use the word collapse but only measurement, or more
generally, complementarity of the principally different
measurements) [6], [7], only a {\it relative} but {\it not an
absolute} phenomenon. Moreover, Bohr compares the relative
character of the collapse within standard quantum mechanical
formalism with the relative character of the space-time in
Einstein's general theory of relativity. In this sense at the end
of his remarkable comment of EPR paradox [6] Bohr observes the
following :

"Especially, the singular position of measuring instruments in the
account of quantum phenomena, just discussed, appears closely
analogous to the well-known necessity in relativity theory of
upholding an ordinary description of all measuring processes,
including a sharp distinction between space and time coordinates,
although the very essence of this theory is the establishment of
new physical laws, in the comprehension of which we must renounce
the customary separation of space and time ideas. The dependence
of the reference system, in relativity theory, of all readings of
scales and clocks may even be compared with essentially
uncontrollable exchange of momentum and energy between the objects
of measurements and all instruments defining the space-time system
of reference, which in quantum theory confront us with the
situation characterized by the notion of complementarity. In fact
this new feature of natural philosophy means a radical revision of
our attitude as regards physical reality, which may be paralleled
with the fundamental modification of all ideas regarding the
absolute character of physical phenomenons, brought about by the
general theory of relativity." (p.701.)

In other words in the general theory of relativity all space-time
referential frames at any space-time point are equivalent or
symmetric and none of these frames can be exactly treated as
unique, i.e. absolute (formal existence of the absolute
referential frame corresponds,in fact,to absolute breaking of the
general relativistic dynamical symmetry), on the one hand. It
means that symmetric general relativistic dynamics, that is, of
course, subluminal, can completely describe all exact general
relativistic phenomenons. On the other hand, at any point of the
Riemannian space-time there is unique local Euclidian referential
frame that can be locally approximately, i.e. effectively treated
as the "absolute" (in this way general relativistic symmetry
becomes locally  approximately, i.e. effectively broken, precisely
hidden). Also, by  a sufficiently small gravitational field and
sufficiently small speeds (ten or many times smaller than the
speed of light), a transition from local Euclidian referential
frame in some close to Euclidian frame  can be effectively
approximated by appearance of the Newtonian inertial forces in
given local Euclidian frame. It denotes the approximate transition
from Einstein's general theory of relativity to Newton's classical
mechanics. But for a sufficiently large gravitational field or
sufficiently large speeds (comparable with the speed of light)
Newton's classical mechanics becomes an incorrect approximation of
the Einstein's general theory of relativity.

Similarly, on the one hand, Bohr suggests that in the standard
quantum mechanical formalism, all bases, representing referential
frames in Hilbert space of the states of quantum system are
equivalent or symmetric and none of these frames can be exactly
treated as unique, i.e. absolute (formal existence of absolute
referential frame corresponds, in fact, to absolute breaking of
the quantum dynamical unitary symmetry,i.e. superposition or  to
absolute collapse). It means that quantum dynamical state
completely (deterministically and objectively) describes the
quantum system in Hilbert space. Also, it means that subluminal
unitary symmetric quantum dynamics (that conserves superposition)
can completely  describe all exact quantum mechanical phenomenons,
including  exact subluminal dynamical interaction between measured
quantum object and measurement  device.On the other hand, it is
possible to apply an approximate, {it hybrid} description of this
interaction between quantum object and measurement device. By such
hybrid description  measurement device is described {\it
approximately, i.e. effectively  classical mechanically}  and
limit of a correct use of this approximation is determined by
Heisenberg's indeterminacy relations. Only in respect to
so-described measurement device, i.e. {\it relatively} quantum
object becomes described {\it effectively quantum mechanically} by
{\it (relatively) collapsed} state from the eigen basis of the
measured observable preferred by given measurement device. In this
way quantum dynamical symmetry, i.e. superposition, becomes {\it
effectively and relatively broken, i.e. effectively hidden}  by
measurement.

More precisely, in Copenhagen interpretation there is a
sufficiently explicit suggestion that a superposition of the
weakly interfering wave packets of the measurement device is
exactly quantum dynamically stable so that it stands exactly
conserved during time. But given superposition is effectively
classical dynamically globally unstable. For this reason only {\it
relatively}, i.e. in respect to effective approximate classical
level of the analysis accuracy this superposition turns
spontaneously (nondynamically and probabilistically) in one
locally classical dynamically stable wave packet with well-known
(Born's) probability. It represents the so-called {\it
self-collapse} on the classically effectively described
measurement device. Simultaneously, on the measured quantum
object, quantum dynamically exactly correlated with measurement
device, {\it relatively},i.e. in respect to selfcollapsed
measurement device, corresponding exact eigen state of the
measured observable appears. It represents {\it relative} collapse
as {\it effective quantum} (but not effective classical)
phenomenon on the measured quantum object. In this way, without
any necessity for an absolute collapse on the classical level of
the analysis or in the macroscopic domains, Copenhagen
interpretation can correctly to reproduce all existing
experimental facts.

It can be added that Bohr himself observed [7] that many of his or
Copenhagen early terms are not quite accurate but that they can be
formalized in a satisfactory way. For example, in early Bohr's
terminology exact conservation of the quantum dynamical unitary
symmetry, or unit norm of the state, is phenomenologically and
simply called "indivisibility of the quant of action" [6],[7],.
Also, Bohr's remarkable term "complementarity" that represents a
conceptual  generalization of the quantum particle-wave duality
principle does not consider any absolute asymmetric duality of the
particle and wave characteristics that exist in de Broglie's,
Bohm'e etc. hidden variables theories [8]. Especially, Heisenberg
[9] cites his discussions with Bohr which many times expressed the
opinion that (particle-wave) duality, even complementarity
represents only a phenomenological expression. It does not belong
to quantum mechanics itself, but it refers to the different
possibilities of the (phase) transitions (that cannot be realized
simultaneously, i.e. in the same conditions) from exact quantum in
the approximate, i.e. effective  classical mechanics. These
complementary transitions occur by different measurements defined
only within the corresponding approximation limits (Heisenberg's
indeterminacy relations). A consequent formalization of Bohr's
relative  collapse concept is given, for example, in [10]. It is
shown that  relative collapse can be modeled by a typical Landau's
continuous phase transition (whose critical values of the order
parameters are determined by Heisenberg's indeterminacy relations)
from exact quantum to the effective, i.e. approximate classical
mechanics.

Really, it is not hard to see that the situation in the quantum
mechanics is conceptually entirely analogous to the situation in
the quantum field theory, especially in the quantum theory of the
electro-weak interactions [11],[12]. Namely, as it is well-known,
the necessity that within a quantum field theory relativistic
locality, i.e. subluminality of the (inter)action is consistently
connected with quantum indeterminacy principle, introduces a
strict condition of the renormalizability of the given quantum
field theory. So, roughly speaking, if the quantum mechanics with
indeterminacy principle would admit a nonrelativistic, i.e.
nonlocal or superluminal completion by hidden variables, then the
corresponding complete quantum field theories cannot be
renoramlizable, or  quantum field theories in the existing form
cannot be possible at all. Renormalizability in the standard model
of the electro-weak interaction is strictly caused by local gauge
symmetry (including its spontaneous (probabilistic) breaking,i.e.
effective hiding) of the quantum electro-weak field dynamics. More
precisely, as it is well-known, the exact local gauge symmetry of
the electro-weak field dynamics does not admit nonzero boson rest
mass. Meanwhile, exactly dynamically stable locally gauge
symmetric electro-weak field is relatively, i.e. in respect to
small perturbation approximation for  low energies, dynamically
nonstable (perturbation series diverges). But, in respect to given
approximation, an electro-weak field  can spontaneously
(nondynamically and probabilistically) turn into one of its
equivalently probable local minimums which generates effectively
nonzero boson mass. However, after this spontaneous local gauge
symmetry breaking (effective hiding), except the three massive
bosons and one masseless photon, there is, according to the
remarkable Goldstone's theorem (that would correspond conceptually
to Bell's theorem)  an additional, fifth, masseless Goldstone's
boson, i.e. "ghost" (spooky or "ethereal") boson. It cannot be
eliminated in any absolute way. Any attempt of the absolute
elimination of such "ghost" boson, that considers explicitly or
implicitly the  absolute existence of such boson before
elimination, leads explicitly or implicitly to the forbidden
absolute (dynamical) breaking of the local gauge symmetry.
Nevertheless it does not mean that in the quantum theory of the
electro-weak interaction there is an absolute asymmetric duality
(material bosons vs. "ghost" boson). Namely, "ghost" bosons can be
correctly treated as a superfluous degree of the freedom only,
i.e. as a possibility for the choice of a Higgs's "rotating"
referential frame. After the choice of given referntial frame,
i.e. after application of so-called Higgs's mechanism, the "ghost"
boson becomes effectively and relatively eliminated without any
exact breaking of the local gauge symmetry of the electro-weak
quantum field dynamics.

Obviously, standard quantum mechanical formalism including its
usual Copenhagen interpretation is deeply conceptually analogous
to practically all modern fundamental physical theories (special
and general theory of relativity and quantum field theory) where
symmetry (including its spontaneous breaking, i.e. effective
hiding by corresponding approximations) represents one of the most
general and significant concepts.

Thus, since collapse is relative, i.e.  principally dependent on
the limits of the approximate classical mechanical description of
the concrete measurement device, within Copenhagen interpretation
of the standard quantum mechanical formalism there is no sense in
discussing the collapse or any corresponding (hidden) variables
before concrete measurement. It represents the main reason why the
premises of EPR or Bell inequality cannot be applied at all within
Copenhagen interpretation of the standard quantum mechanical
formalism or why there are {\it no} superluminal influences here.

In comparison with other interpretations of the quantum mechanics
it can be concluded that Copenhagen interpretation is much more
than a simple,so-called approximationistic interpretation. In such
approximationistic interpretation, eg. [13], roughly speaking, a
measurement device is a quantum mechanically described by a
superposition of the weakly interfering wave packets correlated
with eigen states of the measured observable on the measured
quantum object. But since here an approximate classical level of
the analysis, discretely (for reason of the Heisenberg's
indeterminacy relations) different from the exact quantum level of
the analysis, is not introduced, there is no consistent way of
modeling of the collapse by an individual measurement. Simply
speaking approximationistic interpretations is incomplete, as it
has been pointed out in [14], since it cannot explain  the
empirical fact on the "exact" but not approximate difference
between quantum superposition and quantum mixture on the measured
quantum object after measurement.

Also Copenhagen interpretation is principally different than such
interpretations that can be called super-quantum or macroscopic
hidden variables interpretations. These interpretations or
theories explain collapse as an absolute phenomenon that appears
by an asymmetrically, i.e. nonunitary extended dynamics by
nonlinear terms [15], environment influences [16] etc., in the
macroscopic domains. Nevertheless, since Bell inequality refers to
practically all types of the local hidden variables, Bell
inequality breaking in the real experiments implies that such
macroscopic hidden variables theories must be necessary
superluminal. Moreover, such superluminal macroscopic hidden
variables theories predict not only an incompleteness of the
quantum mechanics but, also, an incompleteness of the classical
mechanics and (special and general) relativistic mechanics in the
macroscopic, nonquantum domains.

It can be added that Copenhagen relative collapse interpretation
is principally different from the famous Everett's relative state
(subsystemic branching), i.e. many world interpretation of the
quantum mechanics [17]. Namely, soon after the appearance of EPR
and significantly before Bell inequality appearance, Furry [18]
observed quite correctly that in contrast to EPR statement quantum
mechanics cannot be both, consistent and incomplete. EPR authors
stated that quantum mechanics cannot differ pure (with
superposition) and mixed (without superposition but with quantum
mixture) state on a quantum supersystem that holds distant
subsystems. But, Furry showed that within the standard quantum
mechanical formalism there is always a clear difference between
pure and mixed state. It refers not only to a simple quantum
system (without subsystems) but also to a quantum supersystem with
near or distant subsystems. For this reason incompleteness of the
quantum mechanics, i.e. existence of a (subsystemic) quantum
mixtures instead of a (supersystemic) quantum superposition,
induces inconsistency of the quantum mechanics too, and vice
versa. Furry's and similar argumentation applied to Everett's
interpretation yields following. Within standard quantum
mechanical formalism supersystemic superposition is inconsistent
with subsystemic branching.Or Everett's interpretation is
inconsistent with standard quantum mechanical formalism and
existing experimental results [4]. This interpretation can be
consistent only with a hidden variables theory but then it must be
necessarily completed by hidden variables that are not given
explicitly in Everett's original work.Meanwhile, in this case
Everett's subsystemic branching and relative states lose
fundamental character and become an auxiliary (even superfluous)
construction.

Finally Copenhagen relative collapse interpretation is different
from von Neumann's, called  orthodox, interpretation of the
quantum mechanics with absolute collapse [5]. It can be observed
and pointed out that there is a very oft opinion, accepted
implicitly by Griffiths too,that orthodox interpretation
represents a new, more accurate version of the old, less accurate
Copenhagen interpretation. But, in fact, such opinion is incorrect
and Copenhagen and orthodox interpretation are principally
different. According to Copenhagen interpretation, as it has been
explained, only unitary symmetric quantum dynamics is completely
exact and universal. Collapse by measurement represents only a
hybrid, effectively classical on the measurement device and
effective and relative quantum on the quantum object, phenomenon.
Simply speaking, Copenhagen interpretation that proposes quantum
mechanics as a theory with spontaneous symmetry breaking (i.e.
effective hiding ) is not old but very modern since it is
comparable with contemporary quantum field theories with
spontaneous symmetry breaking. On the contrary, in the orthodox
interpretation the classical approximation is completely
eliminated from the quantum dynamics. It, seemingly, causes a more
pure, or completely exact theory than this predicted by Copenhagen
interpretation. But in such "more pure" (or "sterile" for the
aspect of Copenhagen interpretation) theory   there is none way
that the collapse be described by quantum dynamics. For this
reason asymmetric (since it breaks unitary symmetry) collapse,
absolutely independent of the unitary symmetric quantum
dynamics,i.e. absolute collapse, must be {\it ad hoc postulated}
without any clear explanation of its origin. This von Neumann's
projection postulate represents {\it an extension} of the standard
quantum mechanical formalism. On the one hand, i.e. {\it
quantitatively or numerically}, von Neuman's projection postulate
corresponds unambiguously to Born's probabilistic postulate by the
getting of the measurements results on the quantum object. On the
other hand, i.e. {\it qualitatively}, von Neumann's projection
postulate states that collapse appears quite exactly and entirely
on the supersystem (measured quantum object + measurement device),
but not only by Copenhagen hybrid description of the same
supersystem. In this way within orthodox interpretation there is
an absolute asymmetrical duality, i.e. the difference between
unitary symmetric quantum dynamics and absolute asymmetric
collapse any of which must be treated as the exact but incomplete.
In this sense quantum mechanics is absolutely exact but
necessarily absolutely incomplete too. Von Neumann suggested
implicitly that absolute asymmetric duality of the quantum
mechanics in the orthodox interpretation represents the result of
a Cartesian like absolute separation of the material world
(described by quantum dynamics) and immaterial, spiritual world of
an Abstract Ego of the human observer (that generate absolute
asymmetric collapse).

EPR authors, that accepted standard quantum mechanical formalism
and absolute collapse postulate, concluded, i.e.  suggested that
quantum mechanics must be incomplete. But in contrast to von
Neumann's orthodox interpretation EPR authors suggest implicitly
that incompleteness of the quantum mechanics can be removed by
some more general dynamical theory called (subquantum) theory of
hidden variables. Word "subquantum" refers to an implicit
supposition that the level of the hidden variables, in some sense
analogous to the quantum field theory level, lies "down" quantum
mechanical level. Implicitly, EPR authors would admit any,
including the above mentioned macroscopic, or superquantum,hidden
variables theory whose level,in some sense analogous to the
classical mechanical level,lies "up" quantum mechanical level.
Nevertheless, any,subquantum or superquantum, but local hidden
variables theory must satisfy Bell inequality. Meanwhile, since
Bell inequality is broken by exact experiments, it means that
none, neither subquantum nor superquantum, local hidden variables
theory can exist. On the other hand acceptance of a nonlocal
hidden variables theory that can to satisfy the existing
experimental results, needs, as it has been discussed, not only a
radical revision of the standard quantum mechanical formalism. It
needs, also, a radical revision (even rejection) of the classical
mechanics,theory of relativity and quantum field theory. It
represents a very implausible demand (an "impossible mission").

Bell [19] suggested a solution of the hidden variables theory
problem by supposition of such hidden variables, called beables,
that, roughly speaking, except their existence, have no other
physical characteristics. Meanwhile, it is not hard to see that
then there is no real physical criterion for differentiation of
locality, i.e. subluminality and nonlocallity, i.e.
superluminality of such "material" beables, or, for
differentiation of such "material" beables and "imaterial"
Abstract Ego of the human observer. It means only that beable
interpretation is equivalently incomplete as well as orthodox
interpretation. In other words, absolute asymmetric duality and
incompleteness problems that appear by extension of the standard
quantum mechanical formalism by  absolute asymmetric collapse are
equivalent to the same problems that appear by extension of the
standard quantum mechanical formalism by hidden variables.

Also, there are opinions [20 ] that absolute asymmetric duality
and incompleteness of the quantum mechanics does not represent
differentiation between onthological, material vs. immaterial,
concepts, but that it represents a differentiation between
onthological and epistemological concepts. In such, so-called
positivistic interpretations of the quantum mechanics onthological
material concepts can be described by unitary symmetric quantum
dynamics, but epistemological concepts (connected with
measurement) stand completely unclear, i.e. quantum dynamically
undescribable,i.e. unexplainable. The difference between
positivistic and orthodox interpretation is based on a supposition
within positivistic interpretation that quantum mechanics (or any
other physical theory) cannot to describe an individual quantum
object or individual measurement but only a statistical quantum
ensemble of the quantum objects and measurement results. However
it is obvious that it represents only a change of the one by the
other ad hoc postulate that is not inherent but that is added to
standard quantum mechanical formalism. This change does not cause
any diminishing of the principal absolute asymmetric duality, i.e.
incompleteness of  the extended quantum mechanics.

All this points clearly that {\it only} usual Copenhagen
interpretation of the quantum mechanics represents an
interpretation with the relative collapse modeled, in full
agreement with general formalism of the spontaneous symmetry
breaking. It, on the one side, {\it without any} "ethereal",i.e.
metaphysical construction, especially {\it without any}
superluminal influences, satisfies {\it all} existing experimental
results. On the other hand,it corresponds deeply conceptually
(with "relativistic" dynamical symmetry and its spontaneous
breaking as the basical concepts) to practically all other
fundamental physical theories, classical mechanics, theory of
relativity and quantum field theory. On the contrary, {\it all the
other} above-mentioned interpretations of the quantum mechanics
are either inconsistent with existing experimental results or hold
explicitly or iimplicitly {\it absolutely unremovable} "ethereal",
i.e. metaphysical constructions, i.e. an {\it absolute asymmetric
duality}. As an especial form (consequence) of this general
absolute asymmetric duality, the duality.i.e. difference between
the subluminal, i.e. dynamical or causal  and superluminal, i.e.
nondynamical or acausal influences appear.

Now we shall analyze a more or less implicit original Griffiths's
attempt to solve the problems of the foundation of the quantum
mechanics. He starts form the view point of the interpretations
with absolute asymmetric duality. But he attempts to prove that
absolute asymmetric duality, especially subluminal, i.e. dynamical
or causal vs. superluminal, i.e. nondynamical or acausal absolute
duality does not characterize only quantum mechanics but that it
characterizes practically all other fundamental physical theories.
If it were be  true, then quantum mechanics with its absolute
asymmetric duality would not represent any exception but only an
especial example of the general principle of the absolute
asymmetric duality in the nature.

Griffiths states the following :

"Ironically, the experimental confirmation of quantum mechanics
came as something of a shock to the scientific community. But not
because it spelled the demise of "realism" - most physicists had
long since adjusted to this (and for those who could not, there
remained the possibility of {\it nonlocal} hidden variables
theories, to which Bell's theorem does not apply). The real shock
was the proof that {\it nature itself is fundamentally nonlocal}.
Nonlocallity, in the form of the instantaneous collapse of the
wave function (and for matter also in the symmetrization
requirement for identical particles) had always been a feature of
the orthodox interpretation, but before Aspect's experiment it was
possible to hope that quantum nonlocality was somehow a
nonphysical artifact of the formalism, with no detectable
consequences. That hope no longer be sustained, and we are obliged
to reexamine our objection to instantaneous action at a distance."
(p.379.)

It can be shortly observed and pointed out that Griffiths,
obviously, neglects principal aspects of Copenhagen interpretation
discussed above. Within Copenhagen interpretation {\it nature
itself is fundamentally local} while seeming impression on the
fundamental nonlocality of the nature is the consequence of an
incorrect postulate on the absoluteness of the collapse. In this
way Griffiths's conclusions can refer, or {\it must be limited},
only to such an interpretation of the quantum mechanics that {\it
ad hoc postulates} absolute asymmetric duality in one or in other
form.

Further,Griffiths states the following :

"Why {\it are} physicists so alarmed at the idea of superluminal
influences? After all, there are many things that travel faster
than light. If a bug flies across the beam of a movie projector,
the speed of its shadow is proportional to the distance to the
screen; in principle, that distance can be as large as you like,
and hence the {\it shadow} can move at arbitrarily high velocity
(Figure A.4.). However, the shadow does not carry any {\it
energy}; nor can transmit any {\it message} from one point to
another on the screen. A person at point ${\it X}$ cannot {\it
cause anything to happen} at point ${\it Y}$ by manipulating the
passing shadows." (p. 380.)

At first sight it represents only a formal analogy between a
classical, strictly nonquantum, superluminality (by motion of the
bug shadow) and quantum (strictly nonclassical) nonlocality (by
Bell inequality breaking). But by a deeper analysis of the cited
(and further) Griffiths's words it can be concluded that Griffiths
points to a principal analogy. Namely Griffiths says :

"Well,let's consider Bell's experiment. Does the measurement of
the electron {\it influence} on the outcome of the positron
measurement? Assuredly it {\it does} - otherwise we cannot account
for correlation of the data. But does the measurement of the
electron {\it cause} a particular outcome for the positron? Not in
any ordinary sense of the word. There is no way the person
monitoring the electron detector could use his measurement to send
a signal to the person at the positron detector, since he does not
control outcome of his own instrument (he cannot {\it make} a
given electron come out spin up, any more than the person at ${\it
X}$ can effect the passing shadow of the bug). It is true that he
can decide {\it whether to make a measurement at all}, but the
positron monitor, having immediate access only to data as his end
of the line, cannot tell whether the electron was measured or not.
For the lists of the data complied at the two ends, considered
separately, are completely random. It is only when we compare the
two lists later that we discover the remarkable correlations."
(p.381.)

Obviously,Griffiths compares {\it very seriously} the observers of
the electron and positron in EPR paradox and the observers of the
bug shadow at ${\it X}$ and ${\it Y}$ point of the screen. By such
comparison the limitation of the separate, or more accurately
subsystemic, measurements on the electron and positron, that
cannot carry energy or information, are {\it seriously analogous}
to the observation of the observers at ${\it X}$ and at  ${\it
Y}$. Let is say that, for example, the bug moves up-down quite
accidentally with the speed that has absolute value $v$ but whose
direction can be changed practically "instanteneously". Let is say
that,in further admirable simplification, this bug can appear
accidentally, during a small time period, either in a point $x$
corresponding to ${\it X}$ or $y$ corresponding to ${\it Y}$. Then
if during given small time interval, the observer at ${\it X}$
observes bug shadow he will know that the observer at ${\it Y}$
observes light point, i.e. shadow absence,and vice versa.

However we can ask again if all Griffiths'es statements, that are
consistent with orthodox interpretation or nonlocal hidden
variables interpretation,are consistent with Copenhagen
interpretation. The answer is again - no, which can be proved by
Bohr's comment of EPR [6], [7]. Simply speaking according to
standard quantum mechanical formalism initial local unitary
symmetric quantum dynamical interaction correlates electron and
positron as the quantum subsystems in the quantum supersystem,
[electron+positron]. Or, given supersystem according to initial
{\it local} quantum dynamics becomes described by pure correlated
quantum state from Hilbert's space of the states of supersystem.
This state of the supersystem that, according to unitary symmetry,
stands correlated even later when immediate dynamical interaction
between distant subsystems  disappear. According to standard
quantum mechanical formalism given supersystemic pure correlated
state must be exactly different from supersystemic mixture of the
noncorrelated states. Simultaneously, it means that given
supersystem {\it cannot be exactly split} into its subsystems even
if they are distant in sense of the usual space of classical
mechanics (Bohr's {\it indivisibility} of the quant of action!).
Namely, fundamental space of the standard quantum mechanical
formalism is Hilbert's space but not the usual classical
mechanical space of the coordinates. Any explicitly or implicitly
opposite statement leads without standard quantum mechanical
formalism. Further when the electron interacts later exactly
quantum dynamically with corresponding measurement device no
absolute collapse on the supersystem appears and for this reason
there is {\it no} absolute superluminal influence of the electron
on the positron. Given exact quantum dynamical interaction
correlates electron+positron supersystem and electron measurement
device in a new broader supersystem, called supersupersystem,
[electron+positron+electron measurement device]. Given
supersupersystem according to {\it local}  quantum dynamical
interaction between electron and its measurement device becomes
described by a new correlated quantum state from Hilbert's space
of the states of supersupersystem. According to standard quantum
mechanical formalism given supersupersystemic pure correlated
state must be exactly different from supersupersystemic mixture of
the noncorrelated states which means that given supersupersystem
{\it cannot be exactly split} into its subsystems, electron,
positron and electron measurement device even if some of these
subsystems  are distant in sense of the usual classical mechanical
space. In this way exact quantum dynamics by initial interaction
between electron and positron or later interaction between
electron, more precisely [electron+positron] and electron
measurement device, yields {\it none} superluminal influences.
Analogously, exact quantum dynamics by initial interaction between
electron and positron or later interaction between positron, more
precisely [electron+positron] and positron measurement device,
yields {\it no} superluminal influences.

Further,by a hybrid description of given supersupersystem in which
on the effectively classically described electron measurement
device a selfcollapse appears on the supersystem relative collapse
appears as an effective quantum phenomena. Relatively collapsed
supersystem becomes effectively exactly described by a mixture of
the noncorrelated quantum states, which admits an effective
separation of the supersystem in the electron and positron.
Obviously, here is nothing superluminal in given selfcollapse and
relative collapse, i.e. by corresponding measurement realized by
electron measurement device. Analogously, there is nothing
superluminal by measurement by positron measurement device and
corresponding selfcollapse and relative collapse. Also, the
comparison of the results of the measurement of the electron
measurement device and positron measurement device points out that
any of given separate, i.e. subsystemic measurements represents
only an incomplete description of corresponding exact description
of the quantum dynamical interactions. In other words, within
Copenhagen interpretation comparison of the separate measurements
does not yield any conclusion on a superluminal influences between
electron and positron. Given comparison, within Copenhagen
interpretation, points out that any of given separate measurement,
i.e. collapse, must be only a relative but not an absolute
phenomenon.

Griffiths ,finally, concludes :

"We are lead, then to distinguish two types of influences: the
"causal" variety, which produce actual changes in some physical
property or receiver, detectable by measurements on that subsystem
alone, and an "ethereal" kind which do not transmit energy or
information, and for which the only evidence is a correlation in
the data taken on the two separate subsystems - a correlation
which by its nature cannot be detected by examing either list
alone. Causal influences {\it cannot} propagate faster than light,
but there is no compelling reason why ethereal ones should not.
These influences associated with collapse of the wave function are
of the latter type, and fact that they "travel" faster than light
may be surprising, but it is not, after all,
catastrophic."(p.381.)

In other words Griffits states that even if absolute asymmetric
quantum duality, dynamical subluminality vs. superluminality by
absolute collapse, seems surprising, it is not catastrophic at all
since quite similar, moreover identical  "catastrophy" exists in
the classical, strictly nonquantum, motion of the bug shadow. By a
simple extension and corresponding examples  it seems that it can
be stated that given absolute asymmetric duality exists in
practically any fundamental physical theory.

But from Copenhagen's view point there is a serious compelling
reason for the rejection of Griffiths's concept of the absolute
asymmetric duality.First of all, from Copenhagen's view point, the
relative character of the collapse within standard quantum
mechanical formalism represents such serious compelling reason.
Moreover, as it has been discussed, necessity that by the
acceptance of the  absolute asymmetric duality practically all
physical theories (classical mechanics, special and general theory
of relativity, quantum mechanics and quantum field theory) must be
radically revised, even completely broken and rejected, represents
a more serious  compelling reason.

All this causes a serious suspicion about the correctness of
Griffiths's example of the classical duality and superluminality
of the bug shadow motion. Really, it is not hard to see that given
superluminality of the bug shadow contradicts sharply to the
remarkable Roemer's determination of the speed of light by real
astronomical facts on the coming late of the first Jupiter moon
shadow. In further text we shall simply prove that this suspicion
is justified, i.e. that Griffiths's example of the classical bug
shadow superluminal motion is definitely incorrect.

We can imagine that in this second we realize a superluminal
influence at  the Sun even if,as it is well-known,light must to
propagate from the Earth to the Sun during, approximately, 8
minutes and 20 seconds. But if we do not suggest any causal
mechanism for demonstration and explanation of this influence this
influence  represents only an unphysical imagination, but not  any
physical "thing that travels faster than light". Meanwhile
Griffiths's example is not based on an imagination since it
proposes  a causal mechanism, i.e. one-to-one correspondence
between initial and final state of the bug shadow. This causal
mechanism is based on the well-known principles of the classical
mechanics and optics, even if, formally, the shadow itself does
not represent a classical mechanical body. It would mean that if
Griffiths's example is correct classical mechanics is not only
incomplete, but that it must be definitely incorrect. In some way,
Griffiths's example in the classical mechanics corresponds to EPR
in the quantum mechanics. All this needs a more accurate analysis,
i.e. formalization which will be given in further work .

So, suppose that a projector is a permanent source of the mutually
decoherent continuous light beams that is placed in the coordinate
beginning $(0,0)$ in $xOy$ plane. In any continuous light beam,
i.e. in any direction on the right hand-side of the coordinate
beginning continuously filled by light (photons), the speed of
light equals $c=3  108 m/s $.

Suppose that the screen is placed parallel to $y-axis$ at the
distance $L$ from the coordinate beginning along $x-axis$
direction.

Suppose that at the initial time moment a small bug is placed at
the point $(l,0)$ at $xOy$ plane for $0<{\it l}<L$.

Finally, suppose that this bug moves with the constant subluminal
speed $v$, parallel to $y-axis$ direction till $({\it l},s)$ point
in $xOy$ plane where it will be stopped. It means that the bug
will appear at $({\it l},s)$ in the time moment
\begin {equation}
    t = \frac {s}{v}
\end {equation}

According to the introduced suppositions (especially suppositions
of the continuity of the light beams and finite value of the speed
of light), the shadow of the bug will appear at $(L,0)$ point of
the screen only after $[0,t1]$ time interval of the initial coming
late,where
\begin {equation}
    t_{1} = \frac {L}{c}
\end {equation}

As it is not hard to see, according to the same suppositions and
(1),(2), the complete time of the motion of the bug shadow from
$(L,0)$ till $(L,S)$, for
\begin {equation}
    \frac {S}{s} =\frac {L}{{\it l}}
\end {equation}
will be
\begin {equation}
     T = t + t_{2} = \frac {s}{v} + \frac {((L-{\it l})^{2}+(S-s)^{2})^{\frac {1}{2}}}{c}
\end {equation}
Here
\begin {equation}
     t_{2}=\frac {((L-{\it l})^{2}+(S-s)^{2})^{\frac {1}{2}}}{c}
\end {equation}
represents length of the time interval of the coming late of the
bug shadow stopping at $({\it l},s)$ in respect to time moment of
the bug stopping at $(L,S)$. It is caused by finite value of the
speed of light.

Now it is obvious that the average speed of the bug shadow equals
\begin {equation}
      v'= \frac {S}{T} = c \frac {S}{c \frac{s}{v} + ((L-{\it l})^{2}+(S-s)^{2})^{\frac {1}{2}}}
\end {equation}
It can be simply proved that following is satisfied
\begin {equation}
     \frac {S}{c \frac {s}{v} + ((L-{\it l})^{2}+(S-s)^{2})^{\frac {1}{2}}} \leq  1
\end {equation}
Namely, (6) can be simply transformed in
\begin {equation}
      (S-c\frac {s}{v})^{2} \leq (L-{\it l})^{2} + (S-s)^{2}
\end {equation}
Since $v$ is subluminal, i.e since
\begin {equation}
   \frac {c}{v}\geq 1
\end {equation}
it follows finally
\begin {equation}
   (S-c\frac {s}{v})^{2} \leq (S-s)^{2} \leq (L-{\it l})^{2} + (S-s)^{2}
\end {equation}
so that (7), (8) are satisfied.

Thus,according to (6), (7) it follows generally
\begin {equation}
   v' \leq c
\end {equation}
which means that the bug shadow moves really with a {\it
subluminal} speed.

But for a relatively small $L$ and $S$  for which
\begin {equation}
   t \gg  t_{1}
\end {equation}
\begin {equation}
   t \gg t_{2}
\end {equation}
$T$ (4) turns approximately in $t$ and $v'$ (6) turns
approximately in
\begin {equation}
   v'\simeq \frac {S}{t}
\end {equation}
which,according to (1),(3) , yields
\begin {equation}
   v' \simeq \frac {L}{{\it l}} v
\end {equation}

Obtained expression points out that since $L$ is greater than
${\it l}$  the bug shadow speed is greater than the bug speed.
Seemingly, according to (15), for a sufficiently large $L$, i.e.
for
\begin {equation}
      L > {\it l}\frac {c}{v}
\end {equation}
v' can be superluminal, which represents the original Griffiths's
argumentation. However, for such a large $L$ expression (15)
becomes an incorrect approximation of (6) while (6) itself,
according to (11), represents even in this case a subluminal
speed.

Suppose now, quite formally, that that expression (15) is always
correct. We shall prove that in this case there is a causal, but
superluminal information transfer by a person, sender of the
information, at ${\it X}=(L,0)$ point and a person, recipient of
the information, at ${\it Y}=(L,S)$ point. Namely, suppose that
there is a third person, mediator ${\it Z}$ close to the bug that
governs bug motion in the following way. It, during the time
interval smaller or equal to $[0,t]$, can either have no influence
on the bug motion or stop the given bug before, even significantly
before it arrives in the final point of its trajectory $({\it
l},s)$. In the first case, during time interval smaller or equal
to $[0,t]$, recipient will observe the appearance of the bug
shadow at ${\it Y}$, while in the second case during the same time
interval he will not observe given shadow. Simply speaking, there
is a possibility of the information transfer from ${\it Z}$ toward
${\it Y}$.

The distance between ${\it X}$ and ${\it Y}$ equals $S$, while the
distance between ${\it X}$ and ${\it Z}$, as it is not hard to
see, is smaller than or equal to $((L-{\it l})^{2}+s^{2})^{\frac
{1}{2}}$. Suppose that ${\it X}$  during the noted time interval
can inform ${\it Z}$, by an additional way (telephone, radio
waves, etc.) with subluminal speed of the information transfer
$v''$, either that bug be not disturbed or that bug be stopped
before it arrives in $({\it l},s)$. This supposition can be
expressed by the following condition
\begin {equation}
    \frac {((L-{\it l})^{2}+s^{2})^{\frac {1}{2}}}{t} \leq v" \leq c
\end {equation}

On the other hand (15),formally treated as an exact expression,
under the condition (16) on  $v'$ superluminality, can be
presented in the following way
\begin {equation}
     v'= \frac {L}{{\it l}}v = \frac {L}{{\it l}}\frac {s}{t}= \frac {Ls}{{\it l}t} > c
\end {equation}

Expressions (16), (17) yield
\begin {equation}
    \frac {{((L-{\it l})^{2}+s^{2})}^{\frac {1}{2}}}{t}  < \frac {Ls}{{\it
    l}{t}}
\end {equation}
or, after simple transformations,
\begin {equation}
     {\it l} ((L-{\it l})^{2}+s^{2})^{\frac {1}{2}} < Ls
\end {equation}
or, finally,
\begin {equation}
     {\it l}^{2} ((L-{\it l})^{2}+s^{2}) < L^{2}s^{2}
\end {equation}

Since condition (16) does not forbid following conditions
\begin {equation}
   L \gg {\it l}
\end {equation}
\begin {equation}
   L \gg s
\end {equation}
it follows that (21) can, for (22),(23), approximated by
\begin {equation}
    {\it l}^{2} < s^{2}
\end {equation}
which yields
\begin {equation}
    {\it l}< s
\end {equation}

Obviously, for an appropriately chosen ${\it l}$ and $s$,
condition (25) can be satisfied without any principal problem.
However, according to  the introduced suppositions,it means that
between ${\it X}$ and ${\it Y}$ a superluminal information
transfer is possible. (It is not hard to see that analogous
conclusion on the possibility of the superluminal communication
between ${\it X}$ and ${\it Y}$ can be satisfied even without
(22), (23) by  the appropriate choice of ${\it l}$, $s$ and $L$,
which will not be discussed in detail.) For this reason, as well
as for the supposed serious analogy between Griffiths's example
and EPR, inequality (25), or generally,(21), can be considered as
the serious analogy of, metaphorically speaking, an anti-Bell
inequality.

In this way it is proved that Griffiths's example of the classical
superluminal motion of a bug shadow is definitely incorrect.
Moreover, it is proved that under a fictitious supposition that
Griffiths's example is correct between the observer at ${\it X}$
and the observer at ${\it Y}$, superluminal and causal information
transfer is possible.

In conclusion we can shortly repeat and point out the following.
Except for a hope of some physicists, there is no compelling
reason to suppose that the absolute asymmetric duality (in the
form of an absolute collapse, superluminal hidden variables, etc.)
exists in the standard quantum mechanical formalism or any other
fundamental physical theory (classical mechanics, theory of
relativity or quantum field theory). On the contrary, there is a
real compelling reason for consideration that standard quantum
mechanical formalism ant its usual, Copenhagen interpretation
yield a complete description of the quantum phenomenons. It is in
full conceptual agreement with the foundation of all other modern
physical theories. In these theories dynamical symmetry principle
(including its spontaneous breaking, i.e. effective hiding in some
of these theories), that "modificates "relativistically"  all
ideas regarding the absolute character of physical phenomena",
represents one of the most significant basic concept.

\vspace {1.5cm}

The authors gratefully appreciate the helpful discussions and
comments provided by Professor Tristan H$\ddot{\rm u}$bsch.
Authors are very grateful to Sanja Pankovi\'c for corrections of
the form of this work.

\newpage

{\large \bf References}

\begin {itemize}

\item [[1]]  D.J.Griffiths,{\it Introduction to Quantum Mechanics}(Prentice Hall Inc.,Englewood Cliffs,New Jersey,1995.)
\item [[2]]  A.Einstein,B.Podolsky,N.Rosen,Phys.Rev.,{\bf 47},(1935.),777.
\item [[3]]  J.S.Bell,Physics,{\bf 1},(1964.),195.
\item [[4]]  A.Aspect,P.Grangier,G.Roger,Phys.Rev.Lett.,{\bf 47},(1981.),460.
\item [[5]] J.von Neumann,{\it Mathematische Grundlagen der Quanten Mechanik} (Springer Verlag,Berlin,1932.)
\item [[6]]  N.Bohr,Phys.Rev.,{\bf 48},(1935.),696.
\item [[7]]  N.Bohr,{\it Atomic Physics and Human Knowledge} (John Wiley,New York,1958.)
\item [[8]]  F.J.Belinfante,{\it A Survay of Hidden Variables Theories} (Pergamon Press,Oxford,1960.)
\item [[9]]  W.Heisenberg,{\it Der Teil und das Ganze} (Ro.Piper Co.,München,1969.)
\item [[10]]  V.Pankovi\'c, T.H$\ddot {u}$bsch,M.Predojevi\'c,M.Krmar,
{\it From Quantum to Classical Dynamics: a Landau Continuous Phase Transition with Spontaneous Superposition Breaking} ,
quant-ph/0409010 v1 1 Sep 2004
\item [[11]]  F.Halzen,A.Martin,{\it Quarks and Leptons: An Introductory Course in Modern Particle Physics} (John Wiley and Sons,New York,1984.)
\item [[12]]  L.H.Ryder,{\it Quantum Field Theory} (Cambridge University Press,Cambridge,1987.)
\item [[13]]  A.Daneri,A.Loinger,G.M.Prosperi,Nucl.Phys.,{\bf 33},(1962.),297.
\item [[14]]  J.S.Bell,Physics World,{\bf 3},(1990.),53.
\item [[15]]  G.C.Ghirardi,A.Rimini,T.Weber,Phys.Rev.D,{\bf 34},(1986.),470.
\item [[16]]  W.H.Zurek,Phys.Rev.D,{\bf 26},(1982.),1862.
\item [[17]]  H.Everett III,Rev.Mod.Phys.,{\bf 29},(1957.),454.
\item [[18]]  W.H.Furry,Phys.Rev.,{\bf 49},(1936.),393.
\item [[19]]  J.S.Bell,{\it Speakable and Unspeakable in Quantum Mechanics} \\(Cambridge University Press,Cambridge,1987.)
\item [[20]]  A.Peres,Am.J.Phys.,{\bf 52},(1984.),644.

\end {itemize}
\end {document}